\documentclass[]{JHEP3}  
\title{Quantum fields, Noether charges and $\kappa$-spacetime symmetries\thanks{To appear in the proceedings of the workshop ``From Quantum to Emergent Gravity: Theory and Phenomenology", Trieste, Italy,  June 11-15 2007.  Based on work done in collaboration with A.~Agostini, G.~Amelino-Camelia, A.~Marcian\'o and R.~Tacchi.}} 
\author{Michele Arzano\\Perimeter Institute for Theoretical Physics\\
        31 Caroline St. N, N2L 2Y5, Waterloo ON, Canada\\
        E-mail: \email{marzano@perimeterinstitute.ca}} 

\abstract{We discuss how the symmetries of $\kappa$-Minkowski non-commutative spacetime can be described by the $\kappa$-Poincar\'e Hopf algebra.  In particular, we focus on a generalization of the Noether analysis in the $\kappa$-deformed framework which allows us to introduce Noether charges associated with translational symmetries and makes possible to overcome the apparent ambiguities that seemed to be associated with the action of such symmetries.  Moving from the classical to the quantum framework, we will first review past results on path integral quantization of a scalar self-interacting field. We will then show how recent work on the symplectic structure of classical $\kappa$-fields suggests a new approach to the canonical quantization of free fields and and present some new results.} 
\begin{document} 

\section{Introduction}
Quantum groups first emerged in the study of certain quantum integrable systems and soon found applications in many areas of theoretical physics and mathematics \cite{Chari}.  The term ``quantum" refers to the fact that the two main classes of quantum groups are obtained by a deformation of Poisson Lie algebras, a procedure which is reminiscent of the Moyal quantization of Poisson manifolds leading from classical mechanics to non-relativistic quantum mechanics.\\  The interest in a possible role of quantum groups for Planck-scale physics began in the late 80's and early 90's \cite{Majid:1988we, Lukierski:1992dt}.  In particular in \cite{Lukierski:1992dt} a ``quantum" deformation of the Poincar\'e algebra with dimensionful deformation parameter $1/\kappa$, with $\kappa$ of the order of the Planck energy, was proposed as a candidate for the description of particle kinematics at the Planck-scale.  The relevance of $\kappa$-Poincar\'e in the study of non-commutative spacetimes became apparent with the work and Majid and Ruegg \cite{Majid-Ruegg} which unveiled the connection between $\kappa$-Poincar\'e in the ``bicrossproduct" basis and $\kappa$-Minkowski non-commutative spacetime.  A new wave of interest toward $\kappa$-deformed symmetries came with the work of Amelino-Camelia in early 2000s  \cite{dsr1}.  In Amelino-Camelia's proposal the mathematical structure of the $\kappa$-Poincar\'e Hopf algebra was used to describe relativistic kinematics with a fundamental, observer independent, length scale $\lambda=1/\kappa$ set by the Planck length $\lambda\equiv L_p\sim 10^{-33} cm$, the threshold at which the familiar concepts of space and time are supposed to break down due to quantum gravity effects.  Soon after \cite{Amelino-Camelia:2003xp} it was argued that $\kappa$-Poincar\'e-type of symmetries should naturally emerge in the low-energy limit of 2+1 dimensional quantum gravity.  Since then a series of related works have appeared in which possible generalizations of \cite{Amelino-Camelia:2003xp} to 3+1 dimension and to include particles have been proposed (see e.g. \cite{qg3D, Freidel:2005me}).  More recently, in \cite{Arzano:2007nx}, it has been shown  that space-time quantum group symmetries, including $\kappa$-Poincar\'e, naturally emerge in the description of the symmetries of quantum fields in the presence of small departures from locality.\\  
The study of classical and quantum fields enjoying $\kappa$-Poincar\'e Hopf algebra symmetries is of fundamental importance per se and constitutes the main tool to gain physical insights on such effective descriptions of flat space limit of quantum gravity.\\
In the next Section, after a brief review of definitions and properties of Hopf algebras, we introduce the $\kappa$-Poincar\'e Hopf algebra and explain its relation with $\kappa$-Minkowski non-commutative spacetime.  In Section 3 we characterize $\kappa$-Poincar\'e symmetries in the context of classical field theory in terms of their associated conserved charges through a generalization of Nother's theorem to the $\kappa$-deformed case.  In Section IV we discuss the path integral and canonical approach to quantization of scalar fields with $\kappa$-deformed symmetries. Section V contains a summary and closing remarks.
\section{Preliminaries: Hopf algebras, $\kappa$-Poincar\'e and $\kappa$-Minkowski}
The ``language" in which quantum groups and quantum deformed algebras are written is that of Hopf algebras.  We start with a flash introduction to these mathematical objects.  Roughly speaking Hopf algebras are generalizations of usual algebras.  A unital, associative {\bf algebra} $(A,m,\eta)$ is a  $\mathbb{C}$-vector space A, together with a multiplication $m$ and a ``unit" map $\eta$
\begin{equation}
m: A\otimes A\rightarrow A;\,\,\, \eta:\mathbb{C}\rightarrow A 
\end{equation}
which satisfy the following properties\footnote{$id$ is the identity map on $A$} ({\it algebra axioms})
\begin{eqnarray*}
m(m\otimes id)&=&m(id \otimes m)~~~~~~~\mathrm{associativity}\\
m(id \otimes\eta)&=&m(\eta\otimes id)=id~~~~~~\mathrm{unit}\, .
\end{eqnarray*}
If one defines two additional maps, the co-product $\Delta$ and the co-unit $\varepsilon$
\begin{equation}
\Delta: A\rightarrow A\otimes A;\,\,\, \varepsilon: A\rightarrow\mathbb{C}  
\end{equation}
satisfying the following properties ({\it co-algebra axioms})
\begin{eqnarray*}
(\Delta\otimes id)\Delta&=(id\otimes \Delta)\Delta~~~~~~~\mathrm{co-associativity}\\
(id\otimes\varepsilon)\Delta&=(\varepsilon\otimes id)\Delta=id~~~~~~~\mathrm{co-unit} 
\end{eqnarray*}
one has a {\bf bialgebra} $(A,m,\eta,\Delta,\varepsilon)$.  The algebra and ``co-algebra" sectors can be connected by an additional map called the {\it antipode}
\begin{equation}
S: A\rightarrow A 
\end{equation}
such that
\begin{equation}
m(S\otimes id)\Delta=m(id\otimes S)\Delta=\eta\circ\varepsilon\, .
\end{equation}
A $\mathbb{C}$-vector space A equipped with the five maps defined above is called a {\bf Hopf algebra} $(A,m,\eta,\Delta,\varepsilon,S)$.  An example of Hopf algebra is given by the universal enveloping (UE) algebra $U(\mathfrak{g})$ of a Lie algebra $\mathfrak{g}$. If $G\in U(\mathfrak{g})$ the co-product, co-unit and antipode are given by 
\begin{eqnarray*}
\Delta(G)&=&G\otimes 1+1\otimes G\\
\varepsilon(G)&=&0\\
S(G)&=&-G\, .
\end{eqnarray*}
The co-product of $U(\mathfrak{g})$ is {\it co-commutative} i.e.
\begin{equation}
\sigma\circ\Delta=\Delta\circ id
\end{equation}
where $\sigma: A\otimes A\rightarrow A\otimes A$ is the "flip" map $\sigma(a\otimes b)=b\otimes a$.  Hopf algebras equipped with a co-commutative co-product are called {\it trivial}.  Quantum deformations of UE algebras lead to {\it non-trivial} Hopf algebras which are one example of {\it quantum groups}.\\
$\kappa$-Poincar\'e has been originally introduced as the contraction of the quantum deformation  of the Anti-de Sitter algebra $U_q(\mathfrak{so}(3,2))$ \cite{Lukierski:1992dt}.  In the so-called ``bicrossproduct basis" \cite{Majid-Ruegg} the co-products for translations $P_0,P_i$, rotations $M_i$ and  boosts $N_i$ are given by
\begin{eqnarray}
\Delta(P_0)&=&P_0\otimes 1+1\otimes P_0\,\,\,\,\,\,\Delta(P_j)=P_j\otimes 1+e^{-P_0/\kappa}\otimes P_j \nonumber \\
\Delta(M_{j})&=&M_{j}\otimes 1+1\otimes M_{j} \nonumber \\
\Delta(N_j)&=&N_j\otimes 1+e^{-P_0/\kappa}\otimes N_j+\frac{\epsilon_{jkl}}{\kappa}P_k\otimes N_l\, . \label{copro}
\end{eqnarray}
The antipodes read
\begin{eqnarray}
S(M_l)&=&-M_l \nonumber\\
S(P_0)&=&-P_0 \nonumber\\
S(P_l)&=&-e^{\frac{P_0}{\kappa}}P_l \nonumber\\
S(N_l)&=&-e^{\frac{P_0}{\kappa}}N_l+\frac{1}{\kappa}\epsilon_{ljk}e^{\frac{P_0}{\kappa}}P_j M_k  \label{anti}\, ,
\end{eqnarray}
while the co-units are trivial
\begin{equation}
\epsilon(P_{\mu})=\epsilon(M_j)=\epsilon(N_k)=0.
\end{equation}
The Hopf algebra multiplication is implicitly defined (via standard ``commutator bracket" construction from an associative algebra) through the commutators
\begin{eqnarray}
\label{commut}
&[P_{0},P_{j}]=0 \qquad [M_j,M_k]=i \epsilon_{jkl}M_l \qquad [M_j,N_k]=i \epsilon_{jkl}N_l  \qquad [N_j,N_k]=i \epsilon_{jkl}M_l  \nonumber\\
&[P_0,N_l]=-iP_l \qquad [P_l,N_j]=-i\delta_{lj}\Big( \frac{\kappa}{2}  \left(1-e^{-\frac{2 P_0}{\kappa}} \right) +\frac{1}{2 \kappa} \vec{P}^2 \Big)+  \frac{i}{\kappa}P_l P_j \nonumber \\
&[P_0,M_k]=0 \qquad [P_j,M_k]=i \epsilon_{jkl}P_l
\end{eqnarray}
The invariant {\it mass} Casimir is given by
\begin{equation}
{\mathcal C}_{\kappa}(P)=\left( 2\kappa \sinh \left( \frac{P_0}{2 \kappa}\right)  \right)^2-\vec{P}^2e^{\frac{P_0}{\kappa}}  
\end{equation}
In the limit $\kappa\rightarrow\infty$ one recovers the {\it trivial} Hopf algebra $U(\mathcal{P})$ associated to the Poincar\'e algebra $\mathcal{P}$.\\  The $\kappa$-Poincar\'e Hopf algebra describes the symmetries of fields living on $\kappa$-Minkowski non-commutative space-time (NCST) in a sense that will be made precise in the next Section.  Here we briefly discuss the connection between $\kappa$-Poincar\'e and $\kappa$-Minkowski.  The latter is a ``Lie-algebra"-type of NCST
\begin{equation}
[x_m,t] = {i \over \kappa} x_m ~,~~~~[x_m, x_l] = 0~,~~~~~l,m=1,2,3\, .
\end{equation}
Intuitively the connection between $\kappa$-Poincar\'e and  $\kappa$-Minkowski space can be seen in a  rather straightforward way.  Consider the set of plane waves $e^{ipx}$ labeled by  $\kappa$-Poincar\'e momenta 
\begin{equation}
P_{\mu}\vartriangleright e^{ipx}=p_{\mu} e^{ipx}\, .
\end{equation}
As complex valued functions the plane waves $e^{ipx}$ form a vector space over $\mathbb{C}$.  In order to define a product ($*$) on such vector space (i.e. introduce an algebra structure) this has to be compatible with the co-product
\begin{equation}
*(\Delta(P_{\mu})\vartriangleright(e^{iq_1 x}\otimes e^{i q_2 x}))\equiv
P_{\mu}\vartriangleright (e^{iq_1 x} * e^{i q_2 x})
\end{equation}
From the LHS above we can write
\begin{equation}
P_{\mu}\vartriangleright (e^{iq_1 x} * e^{i q_2 x})=
(q_1\dot{+}q_2)(e^{iq_1 x} * e^{i q_2 x})
\end{equation}
with $q_1\dot{+}q_2=(q^0_1+q^0_2;\, \vec{q}_1+e^{-q^0_1/\kappa}\vec{q}_2)$ i.e.
\begin{equation}
e^{iq_1 x} * e^{i q_2 x}\equiv e^{(q_1\dot{+}q_2)x}\, .
\end{equation}
Notice how $q_1\dot{+}q_2\neq q_2\dot{+}q_1$, this means that the non-trivial co-product naturally leads to a {\it non-commutative algebra of functions}. Infact elements of such algebra can be seen as {\bf functions of non-commuting coordinates} $\hat{x}$.  Conversely starting from plane waves as functions of non-commuting coordinates, in order to define a multiplication a {\it normal ordering} prescription is needed.  For example we can put all time coordinates to the right
\begin{equation}
:e^{ip\hat{x}}:\equiv e^{i p_m \hat{x}_m}   e^{-i p_0 \hat{x}_0}
\end{equation}
Our wave exponentials having non-commuting objects in their arguments will combine in a non-trivial way, namely according to the Baker-Campbell-Hausdorff (BCH) formula.  One has \cite{AmelinoCamelia:2001fd}
\begin{equation}
(:e^{i q_1 \hat{x}}:) (:e^{i q_2 \hat{x}}:) =
:e^{i (q_1 \dot{+} q_2) \hat{x}}:
\end{equation}
thus obtaining the same results as if one had started with wave exponentials having momenta labels obeying a non-trivial co-product as in (\ref{copro}).\\  A choice of normal ordering is equivalent to a choice of {\bf Weyl map} $\Omega$
which associates functions of non-commuting coordinates to functions of commuting coordinates whose multiplication is given by a non-commutative $*$-product.  For example for the time-to-the right ordering used above one has the following action of the Weyl map on plane waves
\begin{eqnarray}
\Omega(e^{ipx})&=&e^{i p_m \hat{x}_m}   e^{-i p_0 \hat{x}_0}\\
\Omega(e^{ipx})\cdot\Omega(e^{ikx})&=&\Omega(e^{ipx} *e^{ikx})
\end{eqnarray}
Such choice of map is obviously not unique, for example
\begin{eqnarray}
\Omega_s(e^{ipx})&=&e^{-i p_0 \hat{x}_0/2} e^{i p_m \hat{x}_m}   e^{-i p_0 \hat{x}_0/2}\\
\Omega_s(e^{ipx})\cdot\Omega_s(e^{ikx})&=&\Omega_s(e^{ipx} *_s e^{ikx})\, ,
\end{eqnarray}
where the subscript $s$ stands for time-``symmetric" ordering.  These choices of map must correspond to equivalent descriptions of the same field.  In the next Section we will show how this degeneracy in the choice of Weyl maps potentially gives rise to ambiguities in the description of field symmetries and discuss how such difficulties can be overcome.
\section{$\kappa$-symmetries of classical fields and Noether charges}
The description of the action of $\kappa$-Poincar\'e space-time symmetries on fields is greatly simplified by the use of the Weyl maps introduced above.  For our illustrative purposes it will suffice to restrict to plane waves.  It can be shown \cite{Ale2} that, for example, the action of rotations is ``classical" 
\begin{eqnarray*}
M_j\vartriangleright\Omega(e^{ipx})&=&\Omega(M^c_j\vartriangleright e^{ipx})\\
M_j\vartriangleright(\Omega(e^{ipx})\cdot\Omega(e^{ikx}))&=&\Omega(M^c_j\vartriangleright e^{ipx} *e^{ikx}+e^{ipx} *M^c_j\vartriangleright e^{ikx})
\end{eqnarray*}
i.e. the action of $\kappa$-Poincar\'e rotations on a single plane wave corresponds to the action of standard (``classical") rotations ($M^c_j$) while the action on the product of two plane waves is given by the usual Leibnitz rule.  On the other hand boosts have a  ``deformed" action on plane waves
\begin{eqnarray*}
N_j\vartriangleright\Omega(e^{ipx})&=&\Omega(N^{\kappa}_j\vartriangleright e^{ipx})\\
N_j\vartriangleright(\Omega(e^{ipx})\cdot\Omega(e^{ikx}))&=&
\Omega(*(\Delta(N^{\kappa}_j)\vartriangleright (e^{ipx} \otimes e^{ikx})))
\end{eqnarray*}
where the non-trivial action of $N^{\kappa}_j$ comes from the deformed commutator between boosts and momenta in (\ref{commut}) and the deformed Leibnitz rule on products of waves is dictated by the non-trivial co-product $\Delta(N_j)$.  Translations exhibit a behavior which differs from that of rotations and boosts.  In fact the action of translations is classical on a single plane wave 
\begin{equation}
P_{\mu}\vartriangleright \Omega(e^{ipx})=\Omega(P^c_{\mu}\vartriangleright e^{ipx})
\end{equation}
but due to the non-trivial coproduct for the $P_i$'s the action is deformed on products of waves
\begin{equation}
P_i\vartriangleright(\Omega(e^{ipx})\cdot\Omega(e^{ikx}))=
\Omega(P^c_i\vartriangleright e^{ipx} *e^{ikx}+e^{-P^c_0/\kappa}\vartriangleright e^{ipx} *P^c_i\vartriangleright e^{ikx})
\end{equation}
Now we come to the first {\it apparent} ambiguity due to the degeneracy in the choice of Weyl maps.  The  action of a sensible candidate ``physical symmetry" on the fields should not depend on the choice of Weyl map.  Indeed it can be shown \cite{Ale2} that rotations and boosts are not affected by the degeneracy in the choice of Weyl maps, for example for rotations one has
\begin{equation}
M^s_i\vartriangleright\Omega_s(e^{ipx})=
M_i\vartriangleright\Omega(e^{ipx})\, .
\end{equation}
In the case of translations the actions 
\begin{eqnarray*}
P_{\mu}\vartriangleright\Omega(e^{ipx} *e^{ikx})\\
P^s_{\mu}\vartriangleright\Omega_s(e^{ipx} *_s e^{ikx})
\end{eqnarray*}
involve different co-products for the $P_{\mu}$s.  Each different co-product corresponds to a choice of Weyl map.  Different co-products correspond to different "bases" of the $\kappa$-Poincar\'e (Hopf) algebra $(P_{\mu},M_i,N_i)$ and $(P^s_{\mu}, M^s_i, N^s_i)$.  Thus we have that
\begin{equation}
P^s_{\mu}\vartriangleright\Omega_s(e^{ipx})\neq
P_{\mu}\vartriangleright\Omega(e^{ipx})\, .
\end{equation}
It seems that one finds different actions for translation symmetries for each choice of Weyl map, in contrast with the fact that different choices of Weyl maps should correspond to descriptions of the {\it same} physical field.  A solution of this apparent paradox can be found if one takes a more ``pragmatical" approach to the description of symmetries for a given field \cite{knoeth}.  For simplicity we focus on the case of a scalar field.  We characterize a symmetry transformation in terms of the infinitesimal variations 
\begin{eqnarray}
x_{\mu}\rightarrow x'_{\mu}& = & x_{\mu}+dx_{\mu}\\
f(x)\rightarrow f'(x)&=&f(x)+i P_{\mu}f(x)dx_{\mu}=f(x)+df(x)\, .
\end{eqnarray}
Notice that in order to properly define the variations above one needs to specify $dx_{\mu}$s which must obey appropriate commutation relations with the coordinates
\footnote{In dimension $>4$ there exist other choices for the $dx_{\mu}$s see e.g. \cite{Freikono}.} $[x_j +dx_{j},x_0+ dx_0]= \frac{i}{\kappa} (x_j+dx_{j})$ etc.
It seems that things are getting even worse since another factor ordering ambiguity arises in the expression for the differential $df$, namely should one choose $df=iP_{\mu}f(x)dx_{\mu}$ or $df=idx_{\mu}P_{\mu}f(x)$?  It turns out that imposing the Leibnitz rule for such differentials $d(fg)=(df)g+f(dg)$ restricts the choices to one given ordering!  For example the {\it unique choice} for the ``time-to-the-right" Weyl map is
\begin{equation}
df=idx_{\mu}P_{\mu}f(x)\, .
\end{equation}
One can show that a different choice of the Weyl map affects the action of the $P_{\mu}$s but leads to the same $df$ \cite{knoeth}.  This observation is actually the starting point for the derivation of the Noether charges associated with the translational symmetries of a classical field. As an example consider a free massless scalar field $\Phi$ on $\kappa$-Minkowski whose action is given by
\begin{equation}
S[\Phi]=\int d^4x\, \mathcal{L}[\Phi(x)]= \int d^4x {1 \over
2} \tilde{P}_{\mu}\Phi \tilde{P}^{\mu}\Phi
\end{equation}
where 
$\tilde{P}_{0}=\left(2\kappa\right)\sinh({P}_{0}/2\kappa )\,\, ,\tilde{P}_{j}
=e^{{P}_{0}/2\kappa} {P}_{j}$.
Such field obeys a deformed Klein-Gordon equation of motion 
\begin{equation}
\label{kkg}
{\mathcal C}_{\kappa}(P_{\mu})\Phi \equiv  \left[ \left( 2\kappa \right)^2 \sinh^2\left( \frac{P_0}{2\kappa} \right)-e^{P_0/\kappa} \vec{P}^2 \right]\Phi=0\, .
\end{equation}
Making use of the definition of $d\Phi$ under translation given above a Noether analysis is rather straightforward and leads to the following conserved energy-momentum charges (cfr. \cite{knoeth} for details)
\begin{equation}
\label{chrg}
Q_{\mu}=\int d^4p
\, {e^{3 p_0/\kappa} \over 2} p_{\mu} \tilde{\Phi}(p_0, \vec{p})
\tilde{\Phi}(-p_0,- e^{p_0/\kappa}\vec{p})
 \frac{p_0}{|p_0|}\delta({\mathcal C}_{\kappa}(p_{\mu}))
\end{equation}
with $\Phi(\hat{x})=\int d^4q\,\tilde{\Phi}(q)\,\Omega(e^{i q x})$.\\
We would like to point out here that most of the interest in $\kappa$-Minkowski frameworks from a phenomenological point of view has been motivated by the possible deformations of the energy-momentum dispersion relation suggested by the deformed mass Casimir ${\mathcal C}_{\kappa}(p)$.  Indeed it has been suggested that such ``in vacuo dispersion" could lead to observable effects like the explanation of the existence of trans-GZK events in the cosmic ray spectrum or time-of-flight tests using gamma ray bursts (see e.g. \cite{AmelinoCamelia:2004hm} and references therein).  The importance of the result of \cite{knoeth} discussed above is that for the first time we were able to define translations in $\kappa$-Minkowski {\it unambiguously} and the Noether charges (\ref{chrg}) do indeed obey a deformed dispersion relation.\\  What has been said so far concerned the (deformed) symmetries of classical fields. The obvious next step is to try to extend the above consideration to quantum fields.  This turns out to be a non-trivial task and below we report on past and present attempts to the quantization of $\kappa$-fields.
\section{Towards $\kappa$-quantum fields: path integral and canonical quantization}
An early proposal of a scalar quantum field theory with quartic self-interaction based on path integral quantization strategy was proposed in \cite{AmelinoCamelia:2001fd}. The idea was to start from generating functional on $\kappa$-Minkowski
\begin{equation}
Z[J(x)]=\int {\mathcal D}[\phi]\,e^{ i \int
  d^4x \, [\frac{1}{2} \partial^{\mu}\phi(x)\partial_{\mu}\phi(x)
-\frac{m^2}{2} \phi^2(x) - \frac{\lambda}{4!}\phi^4(x)
+\frac{1}{2} J(x)\phi(x)+ \frac{1}{2} \phi(x)J(x)]}~,
\label{zeta1st}
\end{equation}
where a time-to-the-right normal ordering of the non-commuting coordinates is understood.  Using a $\kappa$-Fourier transform one can rewrite the partition function above in energy-momentum space.  Defining an appropriate generalization of the functional derivative
\begin{equation}
\frac{\delta F(f(p))}{\delta f(k)}
=\lim_{\varepsilon\rightarrow 0}\frac{1}{\varepsilon}
\left(F[f(p)+\varepsilon\delta^{(4)}(p\,\dot{+}(\dot{-}k))]-F[f(p)] \right)
~,
\label{functder1}
\end{equation}
\begin{equation}
\frac{\delta F[f(p)]}{\delta f(\dot{-}k)}
=\lim_{\varepsilon\rightarrow 0}\frac{1}{\varepsilon}
\left(F[f(p)+\varepsilon\delta^{(4)}(p\,\dot{+}k)]-F[f(p)] \right)
~.
\label{functder2}
\end{equation}
where $\dot{+}$ and $\dot{-}$ are shorthand notations for the co-product and antipode
\begin{equation}
p_\mu \dot{+} k_\mu \equiv \delta_{\mu,0}(p_0+k_0) + (1-\delta_{\mu,0})
(p_\mu +e^{-p_0/\kappa} k_\mu) ~,
\label{coprod}
\end{equation}
\begin{equation}
(\dot{-} p)_\mu \equiv \delta_{\mu,0}(-p_0) + (1-\delta_{\mu,0})
(-e^{p_0/\kappa} p_\mu) ~,
\label{inverse}
\end{equation}
one obtains
\begin{equation}
\bar{Z}[J(k)]
=e^{i\frac{\lambda}{24}
\int\ \delta^{(4)}\left(\dot{\sum}_{k_1,k_2,k_3,k_4}\right)
\prod_{j=1}^4\frac{d^4k_j}{2\pi}
\xi(k_{j,0})\frac{\delta}{\delta J(\dot{-}k_j)}}
\bar{Z}^0[J(k)]
~,
\label{intpart1}
\end{equation}
where
\begin{equation}\label{}
\xi(k_{j,0}) \equiv
2 \left(1+e^{\frac{-3k_{j,0}}{\kappa}}\right)^{-1} ~,
\end{equation}
and
\begin{equation}
\dot{\sum}_{k_1,k_2,k_3,k_4} \equiv
k_1 \dot{+} k_2 \dot{+} k_3 \dot{+} k_4
~.
\label{sumdot}
\end{equation}
One can perform a perturbative expansion and some interesting results appear already at first order in $\lambda$.  For the the Feynman propagator the topological difference between planar and non-planar tadpole graphs leads to different one-loop corrections
\begin{eqnarray}
G^{(2)}_{\lambda}(p\,,\dot{-}p')^{\mathrm{connected}}_{\mathrm{planar}}&\sim& \frac{\delta^{(4)}(p-p')}{({\mathcal C}_{\kappa}(p)-m^2){(\mathcal C}_{\kappa}(p')-m^2)}\int \frac{d^4q }{{\mathcal C}_{\kappa}(q)-m^2}\\
G^{(2)}_{\lambda}(p\,,\dot{-}p')^{\mathrm{connected}}_{\mathrm{non-planar}}&\sim& \int \frac{d^4q\,\, \delta(p_0-p'_0)\,\,\delta^{(3)}(e^{-p_0/\kappa}\vec{q}-\vec{p}+\vec{q}+e^{- q_{0}/\kappa}\vec{p'})}{({\mathcal C}_{\kappa}(q)-m^2)({\mathcal C}_{\kappa}(p)-m^2){(\mathcal C}_{\kappa}(p')-m^2)}\, ,
\end{eqnarray}
while for the tree-level vertex one notices the emergence of non-trivial ``scattering" kinematics 
\begin{equation}
G^{(4)}_{\lambda}(p_1,p_2,\dot{-}p_3,\dot{-}p_4)^{\mathrm{connected}}\sim \frac{\lambda}{4!}\sum_{\mathcal{P}(\dot{-}p_1,\dot{-}p_2,p_3,p_4)}\left[\delta^{(4)}
(\dot{-}p_1\dot{+}\dot{-}p_2\dot{+}p_3\dot{+}p_4)\right]\, .
\end{equation}
A detailed discussion of these interesting features can be found in \cite{AmelinoCamelia:2001fd}.\\  We see however that with respect to the original goal of describing energy and momentum charges carried by quantum fields a path integral approach is of little help.  A canonical quantization framework would be more appropriate for the characterization of such quantities.  Unfortunately in  $\kappa$-Minkowski space-time, where the time coordinate exhibits non-trivial commutation relations with the spatial coordinates, an extension of the standard ``textbook" approach in terms of equal-time commutation relations between the fields and their canonically conjugate momenta turns out to be rather problematic.  Nevertheless recent work on the symplectic structure of classical $\kappa$-field theories \cite{arma}  suggested an alternative strategy to canonical quantization \cite{Arzano:2007ef}.\\
The idea proposed in \cite{arma} is to study the symmetries of classical $\kappa$-fields borrowing the basic tools of the ``covariant phase space" formalism \cite{Crnkovic:1986ex}.  For a classical undeformed massless scalar field the key point of such formalism is {\it to identify the phase space $\Gamma$ of the theory with the space of solutions of the equation of motion $\mathcal{S}$}
\begin{equation}
\{\phi(x);\pi(x)\}\in\Gamma\longleftrightarrow\Phi\in\mathcal{S}\, .
\end{equation}
On $\mathcal{S}$ one defines \cite{Crnkovic:1986ex} a symplectic 2-form $\omega$ (which on the standard phase space manifold of  classical fields on flat spacetime is given by $\omega=\frac{1}{2}\int_{\Sigma_t}\delta\pi \wedge \delta\phi$) in terms of which the dynamics and the observables of the theory are described.  The strategy adopted in \cite{arma} was to define a Poisson map\footnote{A map between Poisson manifolds is a Poisson map if it preserves the Poisson bracket structure.} $\mathfrak{m}$  
\begin{equation}
\mathfrak{m}:\mathcal{S}\rightarrow\mathcal{S}_{\kappa}
\end{equation}
between the standard space of solutions $\mathcal{S}$ and the space of solutions of the  deformed Klein-Gordon equation (\ref{kkg}) $\mathcal{S}_{\kappa}$ in order to introduce a symplectic structure on $\mathcal{S}_{\kappa}$.  The latter can be used to express the conserved charges associated with the symmetries of the $\kappa$-fields.  In fact in the undeformed case the symplectic structure defines a hermitian inner product on the space of complex solutions 
\begin{equation}
(\Phi_1,\Phi_2)=-2i\,\omega(\Phi^*_1,\,\Phi_2)=\int\frac{d^4 p}{(2\pi)^3}\delta(\mathcal{C}(p))\frac{p_0}{|p_0|}\tilde{\Phi^*}_1{(-p)}\tilde{\Phi}_2(p)\, .
\end{equation}
where the delta function with argument $\mathcal{C}(p)=p^2$ puts the $\tilde{\Phi}$ on-shell.  The Noether charges associated with translation symmetries are then given by $Q_{\mu}=\frac{1}{2}(\Phi, P_{\mu} \vartriangleright \Phi)$.  Using the symplectic structure induced on $\mathcal{S}_{\kappa}$ through the map $\mathfrak{m}$ one easily finds for the charges associated with deformed translations
 \begin{equation}
Q^{\kappa}_{\mu}= \int\frac{d^4p}{2(2\pi)^3}\,\,\delta(\mathcal{C}_{\kappa}(p))\,\,\frac{p_0}{|p_0|}
\,\,p_{\mu}\,\,e^{\frac{3p_0}{\kappa}}\,\tilde{\Phi^*}(\dot{-}p)\tilde{\Phi}(p)
\end{equation}
already found in \cite{knoeth} using a completely different procedure.  This new derivation has the advantage of providing us with an inner product for the space $\mathcal{S}_{\kappa}$
\begin{equation}
(\Phi_1,\Phi_2)_{\kappa}=\int\frac{d^4 p}{(2\pi)^3}\,\,\delta(\mathcal{C}_{\kappa}(p))\,\,\frac{p_0}{|p_0|}
\,\,e^{\frac{3p_0}{\kappa}}\,\tilde{\Phi^*}_1(\dot{-}p)\tilde{\Phi}_2(p)\, .
\end{equation}
The connection with canonical quantization becomes apparent if one notices that in the undeformed case, given an inner product on the complexified space of solutions of the classical equation of motion $\mathcal{S}^{\mathbb{C}}$, there is a standard construction of one-particle Hilbert space $\mathcal{H}$ of the corresponding quantum field theory (see e.g. \cite{Wald:1995yp}).  The map $\mathfrak{m}:\mathcal{S}^{\mathbb{C}}\rightarrow\mathcal{S}^{\mathbb{C}}_{\kappa}$
introduced in \cite{arma} allows an analogous construction for the $\kappa$-deformed one-particle Hilbert space $\mathcal{H}_{\kappa}$.  In the standard case the next step in defining the full kinematical Hilbert space of the theory is to take the direct sum of symmerized (anti-symmetrized) $n$-tensor products of $\mathcal{H}$ for bosons (fermions).  For a scalar field the full Hilbert space of the theory is given by the bosonic Fock space
\begin{equation}
\mathcal{F}_s(\mathcal{H})=\bigoplus_{n=0}^{\infty}S_n\mathcal{H}^n\, ,
\end{equation}
where $S_n=\frac{1}{n!}\sum_{\sigma\in P_n}\sigma $ is the sum of all the possible permutations of $n$ objects and $\mathcal{H}^n$ is the $n$-fold tensor product of one-particle Hilbert spaces $\mathcal{H}^n=\underbrace{\mathcal{H}\otimes\mathcal{H}...\otimes\mathcal{H}}_{n-\mathrm{times}}$.  It turns out that in the $\kappa$-deformed case one cannot proceed in an analogous way to define the Fock space of the theory $\mathcal{F}^{\kappa}_s(\mathcal{H})$.  For example the usual ``symmetrized" two-particle state $\frac{1}{\sqrt{2}}(\phi_{\vec{p}}\otimes \phi_{\vec{q}}+\phi_{\vec{q}}\otimes \phi_{\vec{p}})$, with $\phi_{\vec{p}\,,\vec{q}}\in \mathcal{H}_{\kappa}$, is not an eigenstate of $P_{\mu}$ due to the non-trivial coproduct $\Delta(P_{\mu})$.  The solution to this problem was found in \cite{Arzano:2007ef} where it was shown that a construction of $\mathcal{F}^{\kappa}_s(\mathcal{H})$ consistent with the non-trivial co-algebra structure of $\kappa$-Poincar\'e requires the usual tensor product symmetrization to be combined with
a non-trivial exchange of momentum labels.  Consider for example a an orthonormal basis $\{\phi_{\vec{p}}\}$ of the space of solutions.  The two-particle state $|\,p\dot{+}q>\in\mathcal{F}^{\kappa}_s(\mathcal{H})$ which, as the notation suggests is an eigenstate of the four-momentum operator $P_{\mu}$ with eigenvalue $p\dot{+}q$, is given by
\begin{equation}
\label{ksymm}
1/\sqrt{2}(\phi_{p}\otimes \phi_{q}+\phi_{\tilde{q}}\otimes \phi_{\tilde{p}})
\end{equation}
where
\begin{eqnarray}
\tilde{q} & = &(\tilde{q}_0;\,\vec{q}\,e^{-p_0/\kappa})\nonumber\\
\tilde{p} & = &(\tilde{p}_0;\,\vec{p}\,e^{\tilde{q}_0/\kappa})
\label{bsymm}
\end{eqnarray}
with $\tilde{q}_0=\omega^+(\vec{q}\,e^{-\omega^+(\vec{p})/\kappa})$ and $\tilde{p}_0=
\omega^+(\vec{p}\,e^{\tilde{q}_0/\kappa})$.  As one easily verifies the two couples of momentum labels are such that $\tilde{q}\dot{+}\tilde{p}=p\dot{+}q$.  The two-particle state with momentum eigenvalue $q\dot{+}p\neq p\dot{+}q$ can be obtained with an analogous construction.  This ``deformed" symmetrization can be extended to $n$-particle states and unveils a rich combinatorial structure of the kinematical Fock space $\mathcal{F}^{\kappa}_s(\mathcal{H})$ for $\kappa$-quantum fields.  Multi-particle states belonging to such Fock space are best described in terms of graphs from which one can extract all the information regarding the different momenta labels appearing in the $\kappa$-symmetrized sums analogous to (\ref{ksymm}) (for a complete discussion see \cite{Arzano:2007ef}).  The basic observables of the theory, the field operators $\hat{\Phi}$, can be written down in terms of creation and annihilation operators in a usual fashion.  The classical Noether charges associated with translation symmetries
\begin{equation}
Q_{\mu}=\frac{1}{2}(\Phi, P_{\mu} \vartriangleright \Phi)_{\kappa}\, 
\end{equation}
in the quantum context become observables, i.e. operators on $\mathcal{F}_{\kappa}(\mathcal{H})$
\begin{equation}
\hat{Q}_{\mu}=\frac{1}{2}(\hat{\Phi}, P_{\mu} \vartriangleright \hat{\Phi})_{\kappa}\, .
\end{equation}
Using the explicit expression of $\hat{Q}_{\mu}$ one finds for one and two-particle states
\begin{equation}
\label{1pcharge}
<\, p\, | \hat{Q}_{\mu} |\, p>=\frac{1}{2}(p^+_{\mu}- p^-_{\mu})+Q^{Vac}_{\mu}
\end{equation}
where
\begin{eqnarray}
p_{\mu}^+&=&(\omega^+(\vec{p}), \vec{p})\nonumber\\
p_{\mu}^-&=&(\omega^-(\vec{p}), -\vec{p})\, ,
\end{eqnarray}
with $\omega^{\pm}(\vec{p})$ are the positive/negative roots of the deformed Casimir ${\mathcal C}_{\kappa}(p)$, and
\begin{equation}
<p\dot{+}q\,|\hat{Q}_{\mu}|\,p\dot{+}q>=Q^{Vac}_{\mu}+\left[ \frac{1}{2}\left( p^+_{\mu}+\tilde{q}^+_{\mu} \right)- \frac{1}{2}\left( p^-_{\mu}+\tilde{q}^-_{\mu} \right)  \right]
\end{equation}
in which 
\begin{eqnarray}
\tilde{q}_{\mu}^+=(\omega^+(\vec{q}e^{-p_0^+/\kappa}),\vec{q}e^{-p_0^+/\kappa})\nonumber\\
\tilde{q}_{\mu}^-=(\omega^-(\vec{q}e^{-p_0^+/\kappa}),-\vec{q}e^{-p_0^+/\kappa})\, 
\end{eqnarray}
(completely analogous expressions hold, {\it mutatis mutandis}, for the two-particle state $|\,q\dot{+}p>$).
The term $Q^{Vac}_{\mu}$ is a vacuum energy contribution that remarkably, due to the presence of a natural cut-off $\kappa$ in the theory, does not diverge (see \cite{Arzano:2007ef} for details).
\section{Conclusions}
We presented a brief overview of various advances in the study of classical and quantum field theories with $\kappa$-Poincar\'e symmetries.  First we described how it is possible to give a {\it physical} characterization of $\kappa$-Poincar\'e translations in terms of Noether charges for classical fields.  In this way we showed that it is possible to properly define an action for such symmetries.  Indeed the potential ambiguities that seemed to plague the translation sector of $\kappa$-Poincar\'e emerge when one tries to describe these only in terms of the properties of the algebra of would-be-symmetry generators without taking into account the transformation of the field itself.  Given the important role that energy, momentum and their deformed dispersion relation might have in extracting phenomenological predictions from models with $\kappa$-deformed symmetries, it would be desirable to have a characterization of such charges in the quantum realm.  To this end we reviewed early attempts to a quantization of a self-interacting scalar field using a path integral approach and discussed some relevant results.  Energy and momentum charges carried by quantum fields are best described in a canonical quantization framework.  We showed how recent work on (symplectic) geometrical interpretation of $\kappa$-Noether charges leads to the right direction allowing one to easily define a 
one-particle Hilbert space.  The full Fock space for a free quantum scalar field turns out to have a rather complex structure due to the non-trivial co-product for the translation generators.  Finally we provided explicit expressions for the energy-momentum charges of one and two-particle states belonging to the new $\kappa$-deformed Fock space.\\  The characterization of $\kappa$-Fock space given in \cite{Arzano:2007ef} opens a series of new exciting applications for quantum fields enjoying $\kappa$-Poincar\'e symmetries.  The non-trivial behaviors in the multiparticle sector could have, for example, important consequences for the statistical properties of $\kappa$-quantum fields and lead to interesting results for phenomena involving the presence of a causal horizon e.g. Unruh/Hawking effects.  This issues will be addressed in future studies.

\end{document}